\documentclass[twocolumn,aps,prb,showpacs,floatfix,superscriptaddress]{revtex4}
\usepackage{amsmath}
\usepackage{amsfonts}
\usepackage{amssymb}
\usepackage{bm}
\usepackage{citesort}
\usepackage{graphicx}
\usepackage{subfigure}
\usepackage{epstopdf}
\usepackage{url}

\begin{document}
\title{Lagrangian Structure Functions in Turbulence: A Quantitative
  Comparison between Experiment and Direct Numerical Simulation}

\author{L. Biferale} \affiliation{International Collaboration for
  Turbulence Research} \affiliation{Dip. Fisica and INFN, Universit\`a
  di ``Tor Vergata'' Via della Ricerca Scientifica 1, 00133 Roma,
  Italy.}

\author{E. Bodenschatz} \affiliation{International Collaboration for
  Turbulence Research} \affiliation{Max Planck Institute for Dynamics
  and Self-Organization, Am Fassberg 17, D-37077 Goettingen, Germany}
\affiliation{Laboratory of Atomic and Solid-State Physics, Cornell
  University, Ithaca, NY 14853, USA} \affiliation{Sibley School of
  Mechanical and Aerospace Engineering, Cornell University, Ithaca, NY
  14853, USA} \affiliation{Inst. for Nonlinear Dynamics,
  U. Goettingen, Bunsenstrasse 10, D-37073 Goettingen, Germany}

\author{M. Cencini}\affiliation{International Collaboration for
  Turbulence Research} \affiliation{INFM-CNR, SMC Dipartimento di
  Fisica, Universit\`a di Roma ``La Sapienza'', p.zle A.\ Moro 2,
  00185 Roma, Italy} \affiliation{Istituto dei Sistemi Complessi -
  CNR, via dei Taurini 19, 00185 Roma, Italy}

\author{A. S. Lanotte} \affiliation{International Collaboration for
  Turbulence Research}\affiliation{CNR-ISAC and INFN, Sezione di
  Lecce, Str.\ Prov.\ Lecce-Monteroni, 73100 Lecce, Italy.}

\author{N. T. Ouellette} \affiliation{International Collaboration for
  Turbulence Research} 
\affiliation{Max Planck Institute for Dynamics and Self-Organization,
  Am Fassberg 17, D-37077 Goettingen, Germany} \affiliation{Laboratory
  of Atomic and Solid-State Physics, Cornell University, Ithaca, NY
  14853, USA}\affiliation{Department of
  Physics, Haverford College, Haverford, PA 19041, USA}

\author{F. Toschi} \affiliation{International Collaboration for
  Turbulence Research}\affiliation{Istituto per le Applicazioni del
  Calcolo CNR, Viale del Policlinico 137, 00161 Roma, Italy\\ and
  INFN, Sezione di Ferrara, Via G. Saragat 1, I-44100 Ferrara, Italy.}

\author{H. Xu} \affiliation{International Collaboration for Turbulence
  Research} \affiliation{Max Planck Institute for Dynamics and
  Self-Organization, Am Fassberg 17, D-37077 Goettingen, Germany}
\affiliation{Laboratory of Atomic and Solid-State Physics, Cornell
  University, Ithaca, NY 14853, USA}

\begin{abstract} 
  A detailed comparison between data from experimental measurements
  and numerical simulations of Lagrangian velocity structure functions
  in turbulence is presented. By integrating information from
  experiments and numerics, a quantitative understanding of the
  velocity scaling properties over a wide range of time scales and
  Reynolds numbers is achieved. The local scaling properties of the
  Lagrangian velocity increments for the experimental and numerical
  data are in good quantitative agreement for all time lags. The
  degree of intermittency changes when measured close to the
  Kolmogorov time scales or at larger time lags. This study resolves
  apparent disagreements between experiment and numerics.
\end{abstract}
\maketitle

\section{Introduction}

Understanding the statistical properties of a fully developed
turbulent velocity field from the Lagrangian point of view is a
challenging theoretical and experimental problem.  It is a key
ingredient for the development of stochastic models for turbulent
transport in such diverse contexts as combustion, pollutant
dispersion, cloud formation, and industrial
mixing.\cite{P94,Pope,S01,Y02} Progress has been hindered primarily by
the presence of a wide range of dynamical timescales, an inherent
property of fully developed turbulence. Indeed, for a complete
description of particle statistics, it is necessary to follow their
paths with very fine spatial and temporal resolution, on the order of
the Kolmogorov length and time scales $\eta$ and
$\tau_\eta$. Moreover, the trajectories should be tracked for long
times, order the eddy turnover time $T_L$, requiring access to a vast
experimental measurement region.  The ratio of the above timescales
can be estimated as $T_L/\tau_{\eta} \sim R_\lambda$, and the
microscale Reynolds number $R_\lambda$ ranges from hundreds to
thousands in typical laboratory experiments.  Despite these
difficulties, many experimental and numerical studies of Lagrangian
turbulence have been reported over the years.\cite{YP89, VD97, VSB98,
  OM00, LVCAB01, MMMP01, VLCAB02, SYBVLCB03, PINTON1, mlt2003,
  PINTON2, MCB04, ml2004, BBCDLT04, PK, luthi, leveque, liberzon2,
  BBCLT05, OXBB06b, BBBCCLMT06, X06, YPS06, OXBB06, BEC06, berg,
  liberzon1, AGCBW07, YPKL07, MUELLER06} Here, we present a detailed
comparison between state-of-the-art experimental and numerical studies
of high Reynolds number Lagrangian turbulence. We focus on single
particle statistics, with time lags ranging from smaller than
$\tau_{\eta}$ to order $T_L$. In particular, we study the Lagrangian
Velocity Structure Functions (LVSF), defined as
\begin{equation}
  S_p(\tau) = \langle (\delta_{\tau} v)^p \rangle=
  \langle[v(t+\tau)-v(t)]^p \rangle\,,
\label{eq:dlsf}
\end{equation} 
where $v$ denotes a single velocity component.

In the past, the corresponding Eulerian quantities, \textit{i.e.}~the
moments of the spatial velocity increments, have attracted significant
interest in theory, experiments, and numerical studies (for a review
see Ref.~\onlinecite{frisch}). It is now widely accepted that spatial
velocity fluctuations are intermittent in the inertial range of
scales, for $\eta \ll r \ll L$, $L$ being the largest scale of the
flow. By intermittency we mean anomalous scaling of the moments of the
velocity increments, corresponding to a lack of self-similarity of
their probability density functions (PDFs) at different scales. In an
attempt to explain Eulerian intermittency, many phenomenological
theories have been proposed, either based on stochastic cascade models
(\textit{e.g.}~multifractal
descriptions~\cite{rbm,sheleveque,castaign}), or on closures of the
Navier-Stokes equations.\cite{yakhot} Common to all these models is
the presence of non-trivial physics at the dissipative scale, $r \sim
\eta$, introduced by the complex matching of the wild fluctuations in
the inertial range and the dissipative smoothing mechanism at small
scales.\cite{prlgradients,fv} Numerical and experimental observations
show that clean scaling behavior for the Eulerian structure functions
is found only in a range $10 \eta \leq r \ll L$ (see
Ref.\onlinecite{arneoodo} for a collection of experimental and
numerical results).  For spatial scales $r < 10 \eta$, multiscaling
properties, typical of the intermediate dissipative range, are
observed due to the superposition of inertial range and dissipative
physics.\cite{fv}

Similar questions can be raised in the Lagrangian framework: (i) is
there intermittency in Lagrangian statistics?  (ii) is there a range
of time lags where clean scaling properties (\textit{i.e.}~power law
behavior) can be detected?  (iii) are there signatures of the complex
interplay between inertial and dissipative effects for small time lags
$\tau \sim \mathcal{O}(\tau_{\eta})$?

In this paper we shall address the above questions by comparing
accurate Direct Numerical Simulations (DNS) and laboratory
experiments.  Unlike Eulerian turbulence, the study of which has
attracted experimental, numerical and theoretical efforts since the
last thirty years, Lagrangian studies become available only very
recently mainly due to the severe difficulty of obtaining accurate
experimental and numerical data at sufficiently high Reynolds numbers.
Consequently, the understanding of Lagrangian statistics is still
poor. This explains the absence of consensus on the scaling properties
of the LVSF. In particular, there have been different assessments of
the scaling behavior
\begin{equation} 
S_{p}(\tau) = \langle (\delta_{\tau} v)^p \rangle \sim \tau ^{\xi(p)}\,, 
\end{equation}
mainly due the desire to extract a single number, \textit{i.e.}~the
scaling exponent $\xi(p)$, over a range of time lags.

Measurements using acoustic techniques~\cite{MMMP01,PINTON2} gave the
first values of the exponents $\xi(p)$, measuring scaling properties
in the range $10 \tau_{\eta} < \tau < T_L$. Subsequently, experiments
based on CMOS sensors~\cite{X06,OXBB06} provided access to scaling properties
for shorter time lags, $2\tau_\eta \le \tau \le 6 \tau_{\eta}$,
finding more intermittent values, though compatible with
Ref.~\onlinecite{MMMP01}.  DNS data, obtained at lower Reynolds
number, allowed simultaneous measurements in both of these
ranges.\cite{BBCLT05,BEC06} For $10 \tau_\eta \le \tau \le
50\tau_{\eta}$, scaling exponents were found to be slightly less
intermittent than those measured with the acoustic techniques, though
again compatible within error bars. On the other hand, DNS data~\cite{BEC06,MUELLER06,SCHMIDT} for small time lags, $2 \tau_\eta \le
\tau \le 6 \tau_{\eta}$, agree with scaling exponents measured in
Ref.~\onlinecite{X06}.

The primary goal of this paper is to critically compare
state-of-the-art numerical and experimental data in order to analyze
intermittency at both short and long time lags. This is a necessary
step both to bring Lagrangian turbulence up to the same scientific
standards as Eulerian turbulence and to resolve the conflict between
experiment and simulations (see also
Refs.~\onlinecite{MUELLER06,SCHMIDT,beck}).

\begin{figure}
\begin{center}
  \includegraphics[width=0.47\textwidth]{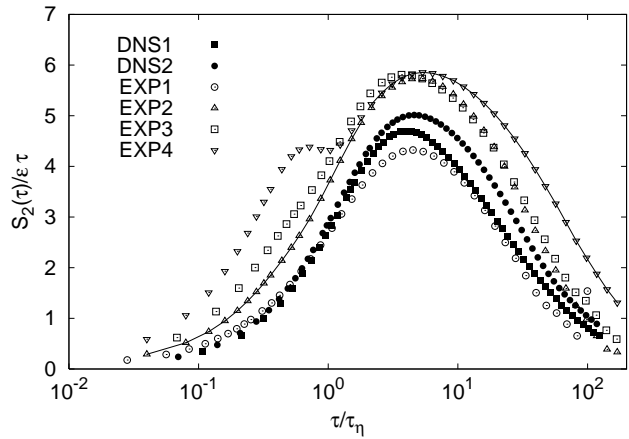}
  \caption{Log-log plot of the second-order LVSF (averaged over the
    three components) normalized with the dimensional prediction,
    \textit{i.e.}  $S_{2}(\tau)/(\epsilon \tau)$, at various Reynolds
    numbers and for all data sets. Details can be found in Tables 1
    and 2. EXP2 and EXP4 refer to the same Reynolds number
    ($R_\lambda=690$), but with different measurement volumes (larger
    in EXP4); in particular EXP2 and EXP4 better resolve the small and
    large time lag ranges, respectively, and intersect for
    $\tau/\tau_\eta\approx 2$. We indicate with a solid line the
    resulting data set made of data from EXP2 (for $\tau/\tau_\eta<2$)
    and EXP4 (for $\tau/\tau_\eta>2$); a good overlap among these data
    is observed in the range $2< \tau/\tau_\eta <8$. For all data
    sets, a extended plateau is absent, indicating that the power law
    regime typical of the inertial range has not yet been achieved,
    even at the highest Reynolds number, $R_\lambda \sim 815 $, in
    experiment.}
\label{fig:0}
\end{center}
\end{figure}

To illustrate some of the difficulties discussed above, in
Fig.~\ref{fig:0} we show a compilation of experimental and numerical
results for the second-order Lagrangian structure function at various
Reynolds numbers (see later for details). The curves are compensated
with the dimensional prediction given by the classical Kolmogorov
dimensional theory in the inertial range~\cite{MY75}, $S_{2}(\tau)=
C_0 \epsilon \tau$, where $\epsilon$ is the turbulent kinetic energy
dissipation. The absence of any extended plateau and the trend with
the Reynolds number indicate that the inertial range, if any, has not
developed yet.  The same trends have been observed in other DNS
studies~\cite{YPS06} and by analyzing the temporal behavior of signals
with a given power-law Fourier spectrum.\cite{LD02}

We stress that assessing the actual scaling behavior of the second
(and higher) order Lagrangian velocity structure functions is crucial
for the development of stochastic models for Lagrangian particle
evolution. Indeed, these models are based on the requirement that the
second-order LVSF scales as $S_2 (\tau) \propto \epsilon \tau $. The issues
of whether the predicted scaling is ever reached and ultimately how
the LVSF deviate as a function of the Reynolds numbers remains to be
clarified.

Moreover, an assessment of the presence of Lagrangian intermittency
calls for more general questions about phenomenological modeling. For
instance, multifractal models derived from Eulerian statistics can be
easily translated to the Lagrangian
framework,\cite{borgas,boffetta,BBCLT05,MMMP01} with some degree of
success.\cite{BBCDLT04,MMMP01,PINTON1}

The material is organized as follows. In Section~\ref{sect:expdns}, we
describe the properties of the experimental setup and the direct
numerical simulations, detailing the limitations in both sets of data.
A comparison of Lagrangian velocity structure functions is considered
in Section~\ref{sec:LVSFcomp}. Section~\ref{sec:lsESS} presents a
detailed scale-by-scale discussion of the local scaling exponents,
which is the central result of the paper. Section~\ref{sec:concl}
draws conclusions and offers perspectives for the future study of
Lagrangian turbulence.

\section{Experiments and Numerical Simulations}
\label{sect:expdns}

Before describing the experimental setup and the DNS we shall briefly
list the possible sources of uncertainties in both experimental and
DNS data.  In general this is not an easy task.  First, it is
important to discern the deterministic from the statistical sources of
errors. Second, we must be able to assess the quantitative importance
of both types of uncertainties on different observables.

{\it Deterministic uncertainties}. For simplicity, we report in this
work the data averaged over all three components of the velocity for
both the experiments and the DNS.  Since neither flows in the
experiments nor the DNS are perfectly isotropic, a part of the
uncertainty in the reported data comes from the anisotropy.  In the
experiments the anisotropy reflects the generation of the flow and the
geometry of the experimental apparatus.  The anisotropy in DNS is
introduced by the finite volume and by the choice of the forcing
mechanism.  In general, the DNS data are quite close to statistical
isotropy, and anisotropy effects are appreciable primarily at large
scales. This is also true for the data from the experiment, especially
at the higher Reynolds numbers. An important limitation of the
experimental data is that the particle trajectories have finite length
due both to finite measurement volumes and to the tracking algorithm,
which primarily affect the data for large time lags. It needs to be
stressed, however, that in the present experimental set up due to the
fact that the flow is not driven by bulk forces, but by viscous and
inertial forces at the blades, the observation volume would anyhow be
limited by the mean velocity and the time it takes for a fluid
particle to return to the driving blades. At the blades the turbulence
is strongly influenced by the driving mechanism. Therefore, in the
experiments reported here the observation volume was selected to be
sufficiently far away from the blades to minimize anisotropy.  For
short time lags, the greatest experimental difficulties come from the
finite spatial resolution of the camera and the optics, the image
acquisition rate, data filtering and post-processing, a step necessary
to reduce noise.  For DNS, typical sources of uncertainty at small
time lags are due to the interpolation of the Eulerian velocity field
to obtain the particle position, the integration scheme used to
calculate trajectories from the Eulerian data, and the numerical
precision of floating point arithmetic.

The {\it statistical uncertainties} for both the experimental and DNS
data arise primarily from the finite number of particle trajectories
and--especially for DNS--from the time duration of the simulations.
We note that this problem is also reflected in a residual, large-scale
anisotropy induced by the non-perfect averaging of the forcing
fluctuations in the few eddy turnover times simulated.  The number of
independent flow realizations can also contribute to the statistical
convergence of the data. While it is common to obtain experimental
measurements separated by many eddy turnover times, typical DNS
results contain data from at most a few statistically independent
realizations.

We stress that, particularly for Lagrangian turbulence, only an
in-depth comparison of experimental and numerical data will allow the
quantitative assessment of uncertainties.  For instance, as we shall
see below, DNS data can be used to investigate some of the
geometrical and statistical effects induced by the experimental
apparatus and measurement technique. This enables us to quantify the
importance of some of the above mentioned sources of uncertainty
directly.  DNS data are, however, limited to smaller Reynolds number
than experiment; therefore only data from experiments can help to
better quantify Reynolds number effects.

\subsection{Experiments} 
\label{sec:exp}

The most comprehensive experimental data of Lagrangian statistics are
obtained by optically tracking passive tracer particles seeded in the
fluid. Images of the tracer particles are analyzed to determine their
motion in the turbulent flow.\cite{VD97,VSB98,OXB06}  Due to the
rapid decrease of the Kolmogorov scale with Reynolds number in typical
laboratory flows, previous experimental measurements were often
limited to small Reynolds numbers.\cite{VD97,OM00} The Kolmogorov
time scale at $R_\lambda \sim \mathcal{O}(10^3)$ in a laboratory water
flow was so far resolved only by using four high speed silicon strip
detectors originally developed for high-energy physics
experiments. \cite{LVCAB01,VLCAB02} The one-dimensional nature of the
silicon strip detector, however, restricted the three dimensional
tracking to a single particle at a time, limiting severely the rate of
data collection.  Recent advances in electronics technology now allow
simultaneous three dimensional measurements of $\mathcal{O}(10^2)$
particles at a time, by using three cameras with two-dimensional CMOS
sensors.  High-resolution Lagrangian velocity statistics at Reynolds
numbers comparable to those measured using silicon strip detectors are
therefore becoming available.\cite{X06}

Lagrangian statistics can also be measured acoustically. The acoustic
technique measures the Doppler frequency shift of ultrasound reflected
from particles in the flow, which is directly proportional to their
velocity.\cite{MMMP01,PINTON2} The size of the particles needed for
signal strength in the acoustic measurements can be significantly
larger than the Kolmogorov scale of the flow. Consequently, the
particles do not follow the motion of fluid particles,\cite{VLCAB02}
and this makes the interpretation of the experimental data more
difficult.\cite{PINTON2}

\begin{table*}[t!]
\begin{ruledtabular}
\begin{tabular}{cccccccccccccc}
  No. & $R_\lambda$ & $v'_{rms}$ & $\varepsilon$ & $\eta$ & $\tau_\eta$ & $T_L$ & $N_f$ & meas. vol. & $\Delta x$ & $N_R$ & $N_{tr}$ \\
  &  & (m/s) & (m$^2$/s$^3$) & ($\mu$m) & (ms) & (s) & (f/$\tau_\eta$) & in ($L^3$) & ($\mu$m/pix) &   & \\ \hline
  EXP1 & 350 & 0.11 & 2.0$\times$10$^{-2}$ & 84 & 7.0 & 0.63 & 35 & 0.4$\times$0.4$\times$0.4 & 50 & 500 & 9.3$\times$10$^5$ \\  
  EXP2 & 690 & 0.42 & 1.2 & 30 & 0.90 & 0.16 & 24 & 0.3$\times$0.3$\times$0.3 & 80 & 480 &
  9.6$\times$10$^5$ \\  
  EXP3 & 815 & 0.59 & 3.0 & 23 & 0.54 & 0.11 & 15 & 0.3$\times$0.3$\times$0.3 & 80 & 500 &
  1.7$\times$10$^6$ \\ 
  EXP4 & 690 & 0.42 & 1.2 & 30 & 0.90 & 0.16 & 24 & 0.7$\times$0.7$\times$0.7 & 200 & 1200 &
  6.0$\times$10$^6$ \\  
\end{tabular}
\end{ruledtabular}
\caption{Parameters of the experiments. Column three gives the  
  value of the root-mean-square velocity fluctuations $v'_{rms}$,  averaged over the three components. The integral length scale
  $L\equiv {v'_{rms}}^3 / \varepsilon =7$cm was determined to be
  independent of Reynolds
  number. $T_L \equiv L/v'_{rms}$ is the eddy turnover time. $N_f$ is the temporal resolution of the
  measurement, in units of frames per $\tau_\eta$. The measurement
  volume was nearly a cube in the center of the tank and its linear
  dimensions
  are  given in units of the integral length scale $L$. $\Delta x$ is
  the spatial discretization of the recording system. The spatial
  uncertainty of the position measurements is roughly 0.1$\Delta
  x$. $N_R$ is the number of independent realizations recorded (see
  text). $N_{tr}$ is the number of Lagrangian trajectories measured.}
\label{table:exp}
\end{table*}

The experimental data here presented are discussed in much detail in
Refs.~\onlinecite{X06,OXBB06}. In the following, we only briefly
recall the main aspects of the experimental technique and data sets,
whose parameters are summarized in Table~\ref{table:exp}.

Turbulence was generated in a swirling water flow between
counter-rotating baffled disks in a cylindrical container. The flow
was seeded with polystyrene particles of size $d_p = 25{\mu}m$ and
density $\rho_p = 1.06$g/cm$^3$ that follow the flow faithfully for
$R_\lambda$ up to 10$^3$.\cite{VLCAB02} The particles were illuminated
by high-power Nd:YAG lasers, and three cameras at different viewing
angles were used to record the motion of the tracer particles in the
center of the apparatus. Images were processed to find particle
positions in three-dimensional physical space; the particles were then
first tracked using a predictive algorithm to obtain the Lagrangian
trajectories.\cite{OXB06} Due to fluctuations in laser intensity, the
uneven sensitivity of the physical pixels in the camera sensor array,
plus electronic and thermal noise, images of particles sometimes
fluctuate and appear to blink. When the image intensity of a particle
was too low, the tracking algorithm lost that particle. Consequently,
the trajectory of that particle was terminated. When the image
intensity is high again, the algorithm started a new trajectory. The
raw trajectories therefore contained many short segments that in
reality belonged to the same trajectory.  It is, however, possible to
connect these segments by applying a predictive algorithm in the
six-dimensional space of coordinates and velocities.\cite{XB07} The
trajectories discussed in this paper were obtained with the latter
method which allows for much longer tracks.

The Lagrangian velocities were calculated by smoothing the measured
positions and subsequently differentiating. A Gaussian filter has been
used to smooth the data. Smoothing and differentiation can be combined
into one convolution operation by integration by parts; the
convolution kernel is simply the derivative of the Gaussian smoothing
filter.\cite{MCB04} The width of the Gaussian kernel was chosen to
remove the noise in position measurements, but not to suppress the
fluctuations, whose characteristic time scale is
$\mathcal{O}(\tau_\eta)$ or above. The velocity statistics have been
found to be insensitive to the width $\sigma$ of the Gaussian filter,
provided it is between $\tau_\eta/6$ and $\tau_\eta/3$ (see also
below). The temporal resolution of the camera system in the
experiments reported here was sufficiently high to ensure that the
fluctuations with time scale greater than $\tau_\eta/6$ were well
resolved.

The uncertainty in position measurement, or the spatial resolution, is
directly proportional to the size of the spatial discretization
determined by the optical magnification and by the size of the pixels
on the CMOS sensor. Larger magnification gives better spatial
resolution but also a smaller measurement volume. Indeed the number of
pixels of the camera sensor array is fixed by the chip-size and, at
higher speeds, by the imaging rate.  The dynamic range of the cameras
is not sufficient to cover the entire range of scales of the
turbulence at the Reynolds numbers of interest. Therefore, two sets of
experiments with different magnifications have been performed. The
former set has high spatial resolution and focuses on the small scale
quantities, though with a relatively small measurement volume
(EXP1,2,3 in Table~\ref{table:exp}). Then, in order to probe longer
times and larger scales, the size of the measurement volume in the
second set of measurements was chosen to be slightly smaller than the
integral scale (EXP4 in Table~\ref{table:exp}).  In this data set,
however, the uncertainty in position was larger and the short-time
statistics were severely affected. As a result, in order to have
experimental data covering a wide range of time lags ($\tau_\eta \le
\tau \le 100 \tau_\eta$) at a given Reynolds number, one needs to
merge data from the two different experiments. This could be done at
$Re_{\lambda}=690$, by using data from the small measurement volume
(EXP2) up to times $\tau \sim (6\div 7) \tau_{\eta}$, and using data
from large measurement volume (EXP4) at larger times. The procedure is
well justified as the two data sets match for intermediate time lags.

One noticeable difference between experiments and numerical
simulations is the number of independent realizations included in the
statistics.  While it is difficult to have many statistically
independent DNS results at one Reynolds number, the experimental data
usually contained ${\cal O}(10^3)$ records separated by a time
interval of about $10^2 T_L$. Each of these records lasted for $(1\div
2)\,T_L$. The variation of the velocity fluctuations calculated from
the statistics of many records is shown in Figure~\ref{fig:2a}. As it is
clear from the figure, the three components do not fluctuate about the
same value, indicating the presence of anisotropy which does not
average away even after many eddy turnover times. These effects are
introduced by the flow generation in the apparatus. In the following
the uncertainties in the data sets due to anisotropy were estimated by
the difference between measurements made on different components of
the velocity field.

\subsection{Direct Numerical Simulations}

Nowadays state-of-the-art numerics~\cite{G02,PK,KAN03,BBCLT05} best
suited for Eulerian statistics is able to reach Taylor scale Reynolds
numbers of the order of $R_{\lambda} \sim 1000$ by using up to
$4096^3$ mesh points.\cite{KAN03} Such extremely high Reynolds number
DNS is, however, limited by the impossibility of integrating the flow
for long time durations, due to the extremely high computational
costs.  In Lagrangian studies it is necessary to highly resolve the
Eulerian velocity field to obtain precise out-of-grid interpolation.
The maximum achievable Reynolds number, on the fastest computers, is
currently limited to $R_{\lambda} \sim 600$ in order to accurately
calculate the particle positions and to achieve sufficiently long
integration times.\cite{Y02,BBCLT05,PK,YPS06}

\begin{figure}[b!] 
\begin{center} 
  \subfigure[]{
  \includegraphics[width=0.47\textwidth]{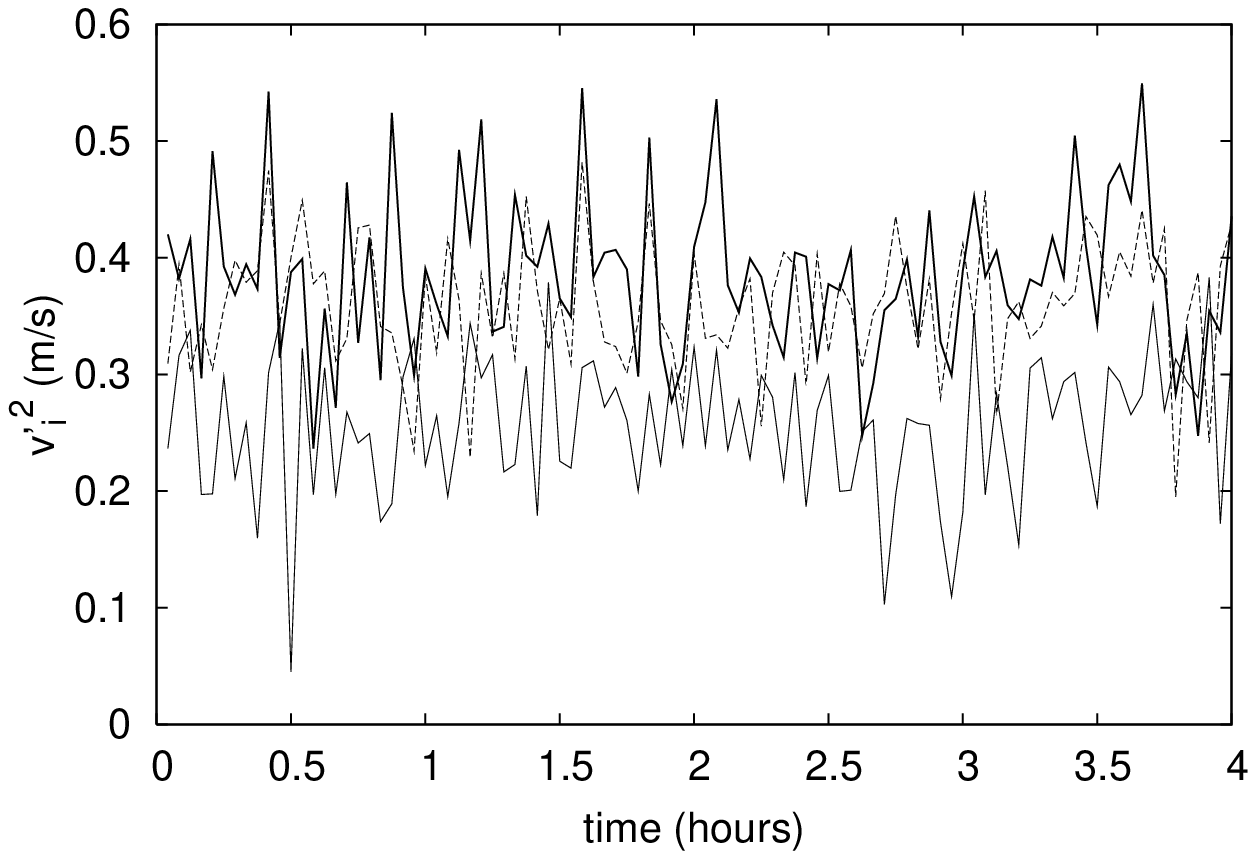}
  \label{fig:2a}
} \hspace{5pt} \subfigure[]{
  \includegraphics[width=0.47\textwidth]{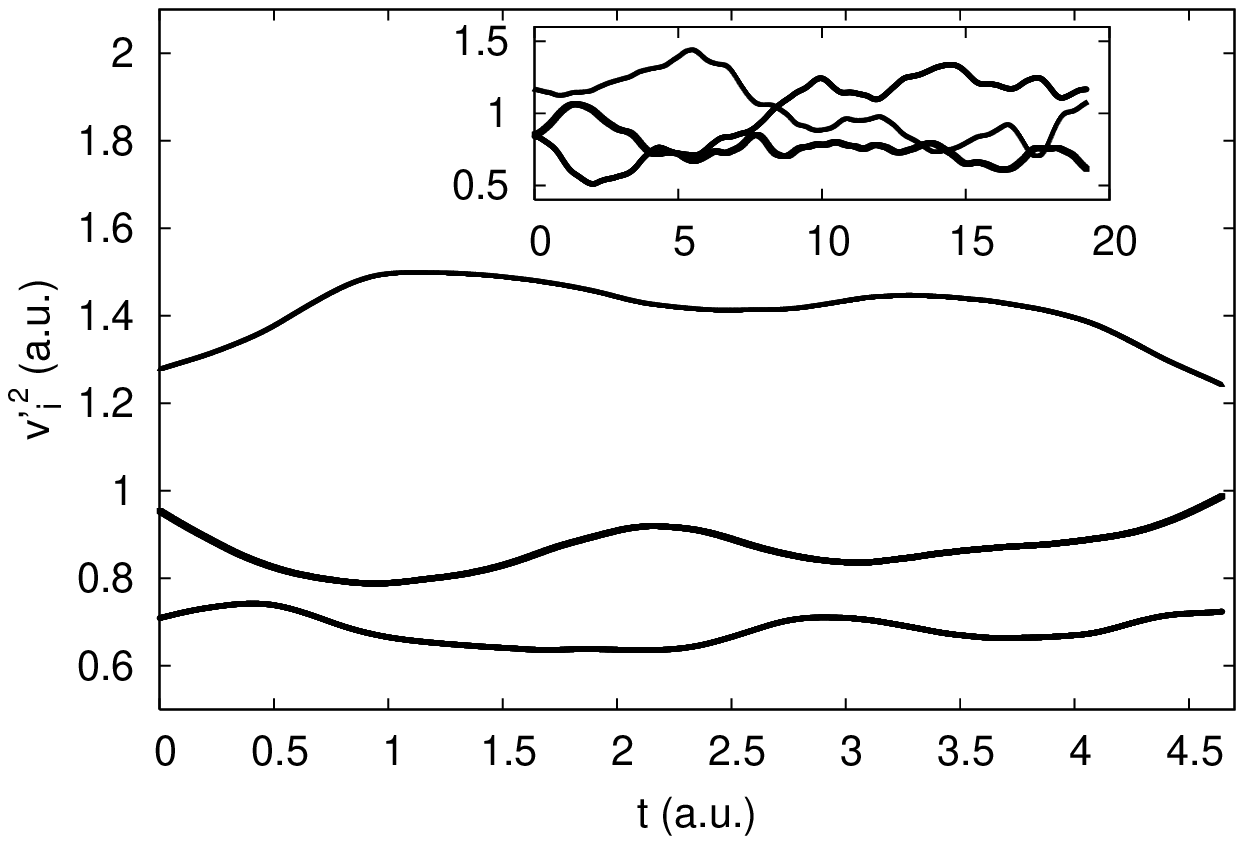}
\label{fig:2b} 
}
\caption{ (b) Time evolution of the components of the velocity
  fluctuation $v'^2_x$ (dashed line), $v'^2_y$ (thick black line) and
  $v'^2_z$ (solid line) for EXP2.  (a) Time evolution of $v'^2_i$,
  with $i=x,y,z$, for DNS2. In the inset we show the same time
  evolution for a DNS at a smaller $R_\lambda\approx 75$ (obtained
  with a spatial resolution of $128^3$ grid points and the same
  forcing), which was integrated for a much longer time. In the latter
  case, the three components fluctuate around the same value, showing
  the recovery of isotropy for long enough time.}
\label{fig:2} 
\end{center}
\end{figure}

Typically, such Lagrangian simulations last for a few large-scale eddy
turnover times, implying some unavoidable remaining anisotropy at
large scales, even for nominally perfectly isotropic forcing. The
simulations analyzed here were forced by fixing the total energy of
the first two Fourier-space shells~\cite{shek}: $E(k_1)= \sum_{|\bm k|
\in I_1} |{\bm \hat v}({\bm k})|$ and $E(k_2)= \sum_{|\bm k| \in I_2}
|{\bm \hat v}({\bm k})|$, where $I_1 = [0.5:1.5]$ and $I_2= [1.5:2.5]$
(the $|\bm k|=0$ mode is fixed to zero to avoid a mean flow).  The
three velocity components can instantaneously be quite different: when
one of the three fluctuates, the others must compensate in order to
keep the total amplitude fixed (see, for instance, Fig.~\ref{fig:2b}
for a visualization of this effect). However, by averaging over many
eddy turn over times - when possible, as for the lower-resolution DNS
shown in the inset of Fig.~\ref{fig:2b}-, the forcing produces a
perfectly statistically isotropic flow. As the remaining large-scale
anisotropy is the main source of uncertainty in the DNS results, we
will estimate confidence intervals from the difference between the
three components.

\begin{table*}[t!]
\begin{ruledtabular}
\begin{tabular}{ccccccccccccc}
  No. & $R_\lambda$ & $v'_{rms}$ & $\varepsilon$ & $\nu$ & $\eta$ & $L$ & $T_L$
  & $\tau_\eta$ & $T$ & $\Delta x$ & $N^3$ & $N_p$ \\  \hline 
  DNS1 &  183 & 1.5 & 0.886 & 0.00205 &0.01 & 3.14 & 2.1 & 0.048 & 5 & 0.012 & $512^3$ &  0.96$\times 10^{6}$ \\ DNS2 &   284 & 1.7 & 0.81 & 0.00088 &0.005& 3.14 & 1.8 &
  0.033 & 4.4 & 0.006 & $1024^3$ & 1.92 $\times10^{6}$ \\ 
\end{tabular}
\end{ruledtabular}
\caption{Parameters of the numerical simulations. Taylor microscale
  Reynolds number $R_\lambda$, root-mean-square velocity fluctuations
  $v'_{rms}$, energy dissipation $\varepsilon$, viscosity $\nu$,
  Kolmogorov length scale $\eta=(\nu^3/\varepsilon)^{1/4}$, integral
  scale $L$, large-eddy turnover time $T_L = L/v'_{rms}$, Kolmogorov
  time scale $\tau_\eta=(\nu/\varepsilon)^{1/2}$, total integration
  time $T$, grid spacing $\Delta x$, resolution $N^3$, and the number
  of Lagrangian tracers $N_p$.}
\label{tab:dns}
\end{table*}

In the simulations, the main systematic error for small time lags
comes from the interpolation of the Eulerian velocity fields needed to
integrate the equation for particle positions,
\begin{equation}
\label{eq:dxdt} 
\dot{{\bm X}}(t) = {\bm v}({\bm X}(t),t)\,.
\end{equation} 
Of course, high-order interpolation schemes such as third-order Taylor
series interpolation or cubic splines partially remove this problem.
Cubic splines give higher interpolation accuracy, but they are more
difficult to use in implementations that rely on secondary
storage.\cite{pk_spline,rovelstad_inter} It has been
reported~\cite{tedeschi} that cubic schemes may resolve the most
intense events better than linear interpolation, especially for
acceleration statistics; the effect, however, appears to be rather
small especially as far as  velocity is concerned.

More crucial than the order of the interpolation scheme is the
resolution of the Eulerian grid in terms of the Kolmogorov length
scale.  To enlarge the inertial range as much as possible, typical
Eulerian simulations tend to poorly resolve the smallest scale
velocity fluctuations by choosing a grid spacing $\Delta x$ larger
then the Kolmogorov scale $\eta$. Since this strategy may be
particularly harmful to Lagrangian analysis, here it has been chosen
to better resolve the smallest fluctuations by choosing $\Delta x
\simeq \eta$ and to use the simple and computationally less expensive
linear interpolation.

We stress that having well resolved dissipative physics for the
Eulerian field is also very important for capturing the formation of
rare structures on a scale $r \approx \eta $. Moreover, as discussed
in Ref.~\onlinecite{yakhot+sreene}, such structures, because of their
filamentary geometry, may influence not only viscous but also inertial
range physics.

Another possible source of error comes from the loss of accuracy in
the integration of Eq.~(\ref{eq:dxdt}) for very small velocities due
to round-off errors. This problem can be overcome by adopting
higher-order schemes for temporal discretization.  For extremely high
Reynolds numbers it may also be necessary to use double precision
arithmetic, while for moderate $R_\lambda$, single precision, which
was adopted in the present DNS, is sufficient for accurate results
(see, \textit{e.g.}, Ref.~\onlinecite{tedeschi}).

Details of the DNS analyzed here can be found
elsewhere~\cite{BBCLT05}; here, we simply state that the Lagrangian
tracers move according to Eq.~(\ref{eq:dxdt}), in a cubic, triply
periodic domain of side $\mathcal{B}=2\pi$. 
DNS parameters are summarized in Table \ref{tab:dns}.

\section{Comparison of Lagrangian Structure Functions}
\label{sec:LVSFcomp}

\begin{figure}[thb]
\begin{center}
  \subfigure[]{
    \includegraphics[width=0.47\textwidth]{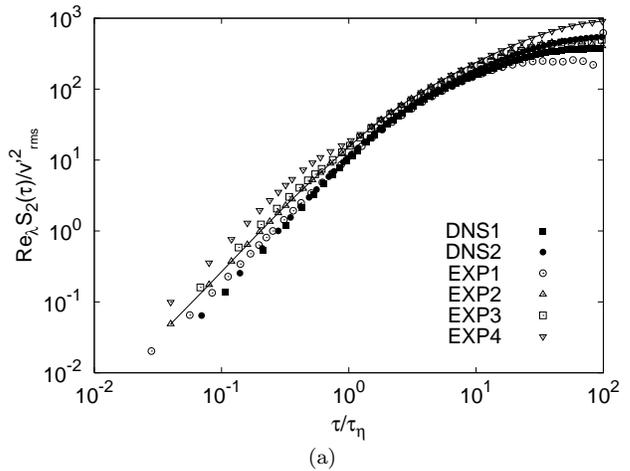}
    \label{fig:LSF2}
  } \hspace{5pt} \subfigure[]{
    \includegraphics[width=0.47\textwidth]{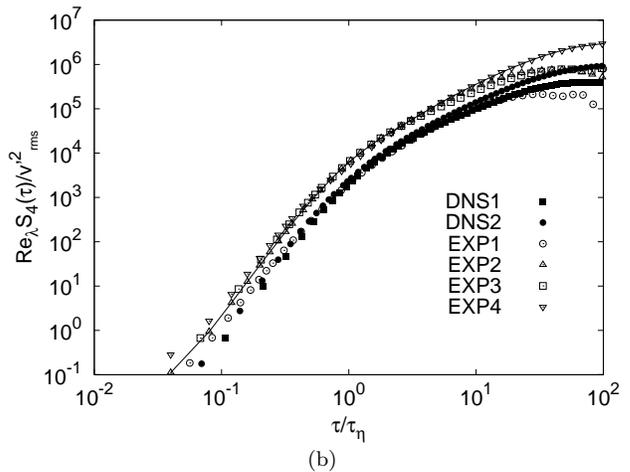}
\label{fig:LSF4}
}
\caption{(a) Log-log plot of the second-order structure function
  compensated as $R_{\lambda}S_{2}(\tau)/{v'_{rms}}^2$ vs
  $\tau/\tau_{\eta}$ for all data sets, at several Reynolds
  numbers. (b) The same for the fourth-order structure function
  $R^2_{\lambda}S_{4}(\tau)/{v'_{rms}}^4$. The solid line is made to
  guide the eye through the two data sets (EXP2 and EXP4) obtained at
  the same Reynolds number in two different measurement volumes, as
  explained in Sect.~\ref{sec:exp}.}
\label{fig:LSF}
\end{center}
\end{figure}

Let us now compare the experimental and numerical measurements of the
Lagrangian velocity structure functions directly.
Figures~\ref{fig:LSF2} and \ref{fig:LSF4} show a direct comparison of
LVSFs of order $p=2$ and $p=4$ for all data sets.  The curves are
plotted using the dimensional normalization, assuming that $S_2(\tau)
= C_0 \epsilon \tau \propto {v'}_{rms}^2 R^{-1}_\lambda
(\tau/\tau_\eta)$ (where we use $\epsilon \approx v'^3_{rms}/L$ and
$T_L/\tau_\eta \propto R_\lambda$).  Such a rescaling can be
generalized as $S_p(\tau) \propto {v'}_{rms}^p R^{-p/2}_\lambda
(\tau/\tau_\eta)^{p/2}$.  Both the $2^{\textrm{nd}}$ and
$4^{\textrm{th}}$ order moments show a fairly good collapse,
especially in the range of intermediate time lags. However, some
dependence can be observed both on $R_\lambda$ (see
Fig.~\ref{fig:LSF4}) and on the size of the measurement volume
(compare EXP2 and EXP4). Both effects call for a more quantitative
understanding.

\subsection{Local Scaling Exponents}
\label{sec:lsESS}

A common way to assess how the statistical properties change for
varying time lags is to look at dimensionless quantities such as the
generalized flatness
\begin{equation} 
F_{2p}(\tau) = \frac{S_{2p}(\tau)}{[S_{2}(\tau)]^{p}}\,.  
\label{eq:flat}
\end{equation} 
We speak of {\it intermittency} when such a function changes its
behavior as a function of $\tau$: this is equivalent to the PDF of the
velocity fluctuations $\delta_{\tau}v$, normalized to unit variance,
changing shape for different $\tau$.\cite{frisch}

\begin{figure}[thb]
\begin{center}
  \includegraphics[width=0.47\textwidth]{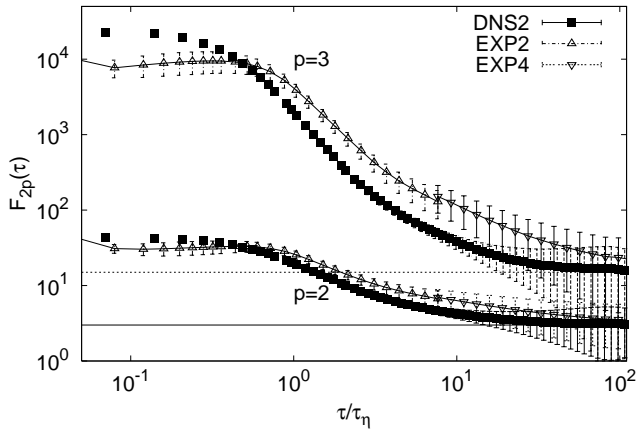}
  \caption{Generalized flatness $F_{2p}(\tau)$ of order $p=2$ and
    $p=3$, measured from DNS2, EXP2, and EXP4. Data from EXP2 and EXP4
    are connected by a continuous line.  The Gaussian values are given
    by the two horizontal lines. The curves have been averaged over
    the three velocity components and the error bars are computed from
    the scatter between the three different components as a measure of
    the effect of anisotropy. Statistical errors due to the limitation
    in the statistics are evaluated by dividing the whole data sets in
    sub samples and comparing the results. These statistical errors
    are always smaller than those estimated from the residual
    anisotropy.}
  \label{fig:flat}
\end{center}
\end{figure}

When the generalized flatness varies with $\tau$ as a power law,
$F_{2p}(\tau) \sim \tau^{\chi(2p)}$, the scaling laws are {\it
intermittent}. Such behavior is very difficult to assess
quantitatively, since many decades of scaling are typically needed to
remove the effects of sub-leading contributions (for instance, it is
known that Eulerian scaling may be strongly affected by slowly
decaying anisotropic fluctuations~\cite{bp}).

We are interested in quantifying the degree of intermittency at
changing $\tau$.  In Fig.~\ref{fig:flat}, we plot the generalized
flatness $F_{2p}(\tau)$ for $p=2$ and $p=3$ for the data sets DNS2,
EXP2 and EXP4.  Numerical and experimental results are very close, and
clearly show that the intermittency changes considerably going from
low to high $\tau$.

The difficulty in trying to characterize these changes quantitatively
is that, as shown by Fig.~\ref{fig:flat}, one needs to capture
variations over many orders of magnitude.  For this reason, we prefer
to look at observables that remain ${\cal O}(1)$ over the entire range
of scales and which convey information about intermittency {\it
  without having to fit any scaling exponent}.  With this aim, we
measured the logarithmic derivative (also called local slope or local
exponent) of structure function of order $p$, $S_{p}(\tau)$, with
respect to a reference structure function,\cite{ESS} for which we
chose the second-order $S_{2}(\tau)$:
\begin{equation}
  \zeta_{p}(\tau) =
  \frac{{\rm d}\log{(S_{p}(\tau))}}{{\rm d}\log{(S_{2}(\tau))}}\,.
\end{equation}
We stress the importance of taking the derivative with respect to a
given moment: this is a direct way of looking at intermittency with no
need of \textit{ad hoc} fitting procedures and no request of power
law behavior. This procedure,\cite{ESS} which goes under the name of Extended
Self Similarity~\cite{ESS} (ESS), is particularly important when
assessing the statistical properties at Reynolds numbers not too high
and/or close to the viscous dissipative range.

\begin{figure*}[t!]
\begin{center}
  \subfigure[]{
    \includegraphics[width=0.7\textwidth]{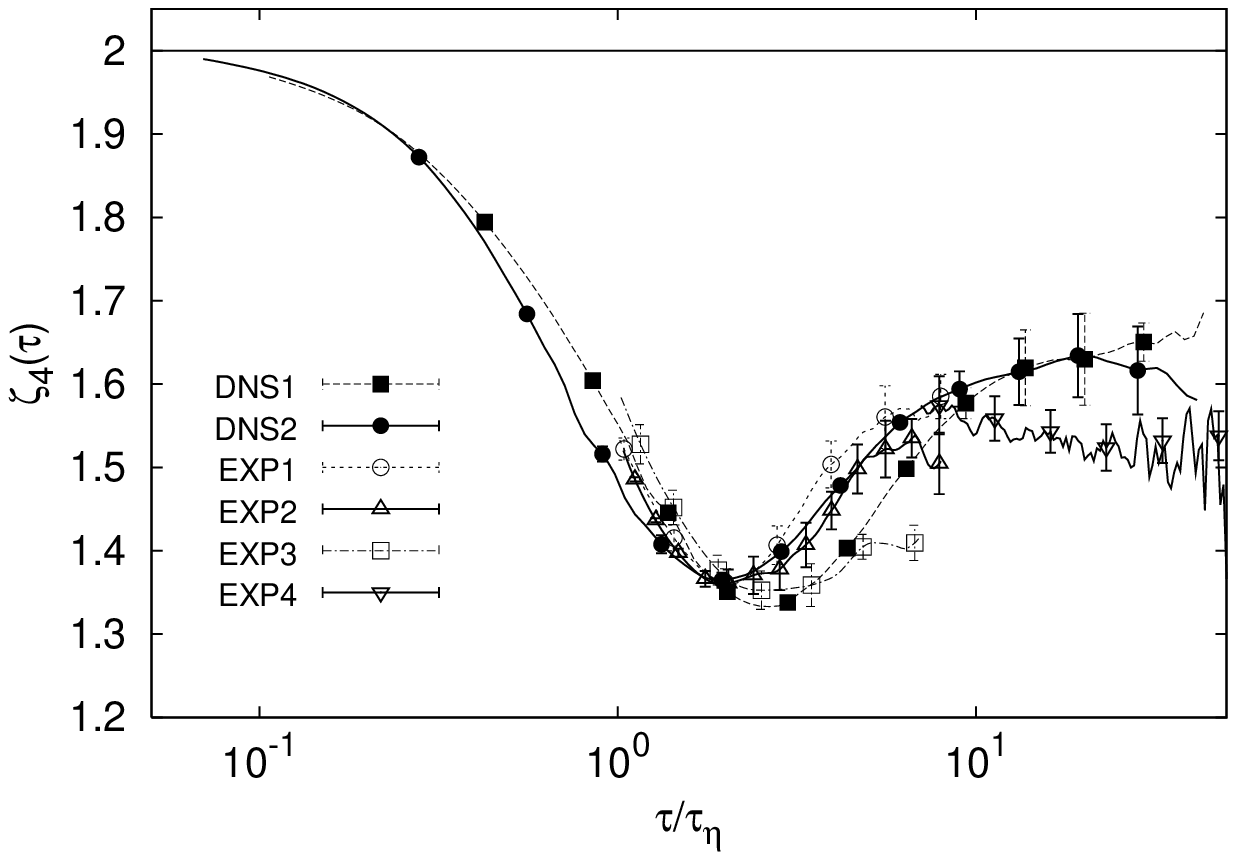}
\label{fig:ess4}
} \subfigure[]{
  \includegraphics[width=0.7\textwidth]{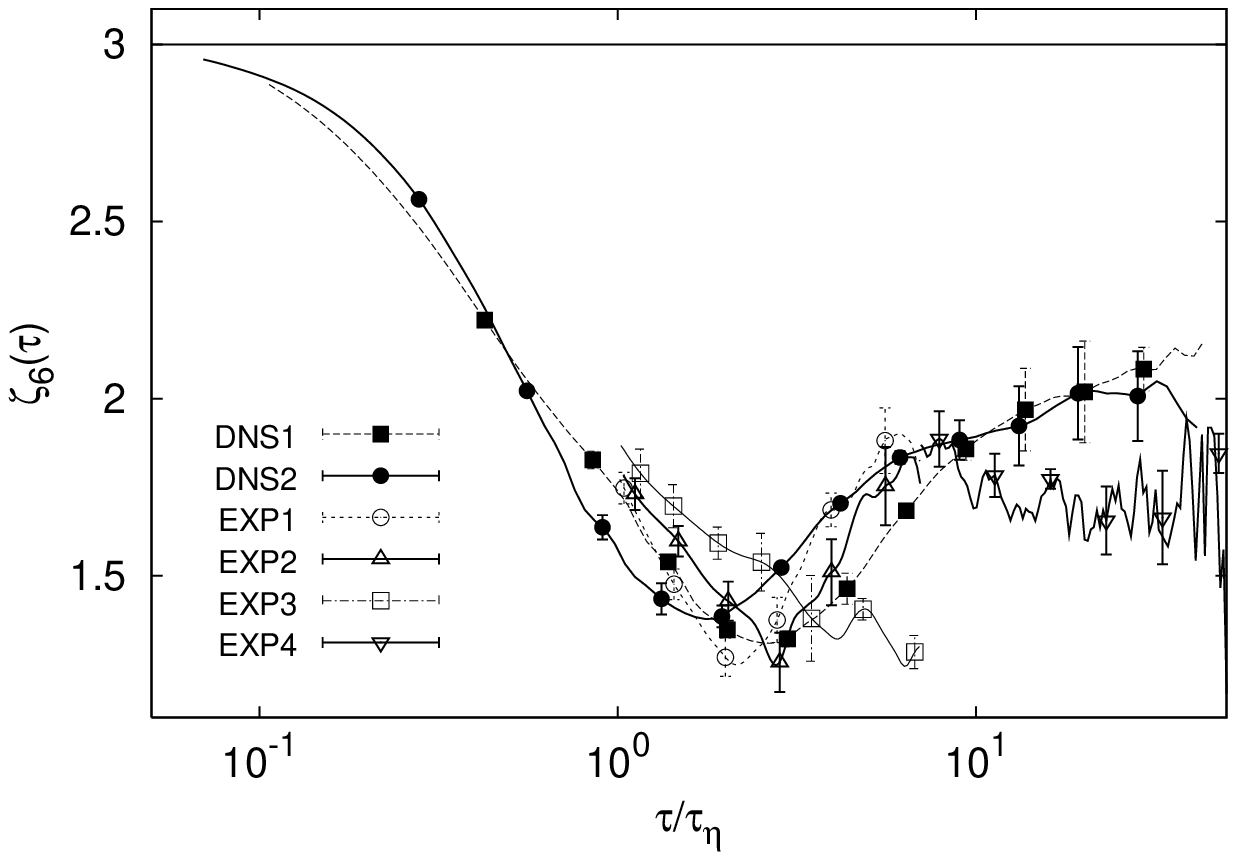}
  \label{fig:ess6}
}
\caption{Logarithmic derivatives $\zeta_p(\tau)$ of structure
  functions $S_p(\tau)$ with respect to $S_2(\tau)$ for orders $p=4$
  (a) and $p=6$ (b). The curves are averaged over the three velocity
  components and the error bars are computed from the statistical
  (anisotropic) fluctuations between LVSFs of different
  components. The horizontal lines are the non-intermittent values for
  the logarithmic local slopes, \textit{i.e.}~$\zeta_p=p/2$. We stress
  that the curves for EXP1,2,3 are shown in the time range $1 \le
  \tau/\tau_{\eta} \le 7$, while the curves for EXP4 (large
  measurement volume) are shown in the time range $7 \le
  \tau/\tau_{\eta} \le 50$.}
\label{fig:6}
\end{center}
\end{figure*}

A non-intermittent behavior would corresponds to
$\zeta_{p}(\tau)=p/2$. In the range of $\tau$ for which the exponents
$\zeta_{p}(\tau)$ are different from the dimensional values $p/2$,
structure functions are intermittent and correspondingly the
normalized PDFs of $\left(\delta_{\tau} v/ \langle(
\delta_{\tau}v)^2\rangle^{1/2}\right)$ change shape with $\tau$.
Figures~\ref{fig:ess4} and \ref{fig:ess6} show the logarithmic local
slopes of the numerical and experimental data sets for several
Reynolds numbers for $p=4$ and $p=6$ versus time normalized to the
Kolmogorov scale, $\tau/\tau_\eta$.  These are the main results of our
analysis.

The first observation is that for both orders $p=4$ and $p=6$, the
local slopes $\zeta_p(\tau)$ deviate strongly from their
non-intermittent values $\zeta_{4}=2$ and $\zeta_{6}=3$.  There is a
tendency toward the differentiable non-intermittent limit $\zeta_p =
p/2$ only for very small time lags $\tau \ll \tau_\eta$.

In the following, we shall discuss in detail the small and large time
lag behavior.

{\it Small time lags.} For the structure function of order $p=4$
(Fig.~\ref{fig:ess4}), we observe the strongest deviation from the
non-intermittent value in the range of time $2 \tau_\eta \le \tau \le
6\tau_{\eta}$.  It has previously been proposed that this deviation is
associated with particle trapping in vortex filaments.\cite{BBCLT05}
This fact has been supported by DNS investigations of inertial
particles.\cite{mlt2003,ml2004,BEC06} The agreement between the DNS
and the experimental data in this range is remarkable. For $p=6$
(Fig.~\ref{fig:ess6}), the scatter among the data is higher due to the
fact that, with increasing order of the moments, inaccuracies in the
data become more important. Still, the agreement between DNS and the
experimental data is excellent. Differently from the $p=4$ case, a
dependence of mean quantities on the Reynolds number is here
detectable, though it lies within the error-bars. The experimental
data set for $p=6$, at the highest Reynolds number ($R_{\lambda} =
815$), show a detectable trend in the local slope toward less
intermittent values in the dip region, $2 \le \tau/\tau_{\eta} \le 6$.
This change may potentially be the signature of vortex destabilization
at high Reynolds number -- which would reduce the effect of vortex
trapping. It is more likely, however, that at this very high Reynolds
number both spatial and temporal resolution of the measurement system
may not have been sufficient to resolve the actual trajectories of
intense events.\cite{BBCLT05} We consider this to be an important open
question for future studies.

{\it Larger time lags.} For $\tau > (6 \div 7) \tau_{\eta}$ up to $T_L$,
the experimental data obtained in small measurement volumes
(EXP1,2,3), are not resolving the physics, as they develop both strong
oscillations and a common trend toward smaller and smaller values for
the local slopes for increasing $\tau$. This may be attributed to
finite volume corrections (see also Sect.~\ref{sec:vol}).  For these
reasons, the data of EXP1,2,3 are not shown for these time ranges.  On
the other hand, the data from EXP4, obtained from a larger measurement
volume, allow us to compare experiment and simulation.  Here the
local slope of the experimental data changes slower very much akin to
the simulations.  This suggests that in this region high Reynolds
number turbulence may show a plateau, although the current data can
not give a definitive answer to this question. For $p=6$, a similar
trend is detected, though with larger uncertainties. The excellent
quantitative agreement between DNS and the experimental data gives us
high confidence into the local slope behavior as a function of time
lag.

In light of these results, we can finally clarify the recent apparent
discrepancy between measured scaling exponents of the LVSFs in
experiments~\cite{X06} and DNS,\cite{BBCLT05} which have lead to some
controversy in the literature.\cite{SCHMIDT,MUELLER06,beck}  In the
experimental work~\cite{X06}, scaling exponents were measured by
fitting the curves in Fig.~\ref{fig:6} in the range $2 \tau_\eta \le
\tau \le 6 \tau_{\eta}$, where the compensated second order velocity
structure functions reach a maximum, as shown in Fig.~\ref{fig:0}
(measuring the fourth and sixth order scaling exponents
$\zeta_p(\tau)$ to be $1.4 \pm 0.1$ and $1.6 \pm 0.1$, respectively).
On the other hand, in the simulations~\cite{BBCLT05} scaling exponents
were measured in the regions in the range of times $10 \tau_\eta \le \tau \le
50 \tau_{\eta}$ (finding the values $\zeta_4= 1.6 \pm 0.1$ and
$\zeta_6= 2. \pm 0.1$).

It needs to be emphasized, however, that the limits induced by the
finiteness of volume and of the inertial range extension in both DNS
and experimental data do not allow for making a definitive statement
about the behavior in the region $\tau > 10 \tau_{\eta}$.  We may ask
instead if the relative extension of the interval where we see the
large dip at $\tau \sim 2 \tau_{\eta}$ and the possible plateau,
observed for $\tau > 10 \tau_{\eta}$ both in the numerical and
experimental data (see EXP4 data set), becomes larger or smaller at
increasing the Reynolds number.\cite{YPKL07} If the dip region --the
one presumably affected by vortex filaments-- flattens, it would give
the asymptotically stable scaling properties of Lagrangian turbulence.
If instead the apparent plateau region, at large times, increases
in size while the effect of high intensity vortex remains limited to
time lags around $(2\div 6) \tau_{\eta}$, the plateau region would
give the asymptotic scaling properties of Lagrangian turbulence. This
point remains a very important question for the future because, as of
today, it can not be answered conclusively neither by experiments nor
by simulations.

\subsubsection{Finite volume effects at large time lags}
\label{sec:vol}

\begin{figure}[t!]
\begin{center}
\includegraphics[width=0.47\textwidth]{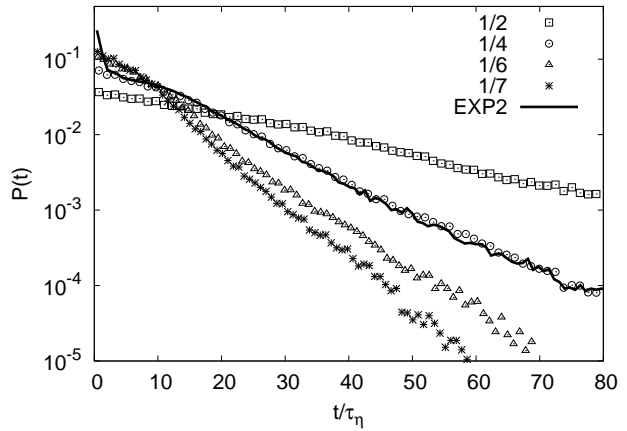}
\caption{Comparison of the probability ${\cal P}(t)$ that a trajectory
  lasts a time $t$ vs $t/\tau_\eta$ for the experiment EXP2 and for
  DNS2 trajectories in different numerical measurement domains ${\cal
    L}/\mathcal{B}=1,1/2,1/4,1/6,1/7$. }
\label{fig:times}
\end{center}
\end{figure}

\begin{figure}[t!]
\begin{center}
\subfigure[]{
  \includegraphics[width=0.47\textwidth]{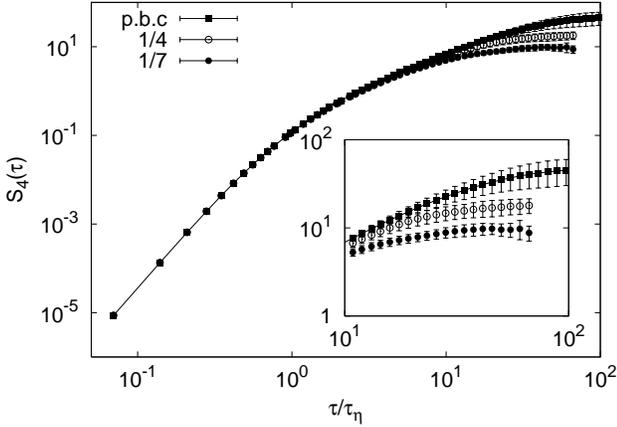}
\label{fig:aleA}
} \subfigure[]{
  \includegraphics[width=0.47\textwidth]{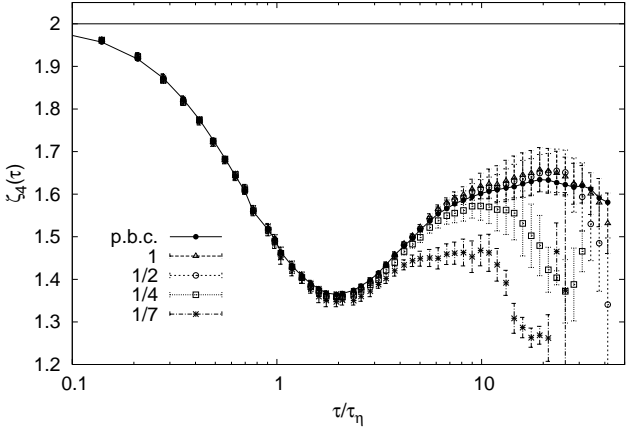}
\label{fig:aleB}
}
\caption{(a) The fourth-order structure function $S_4(\tau)$ vs
  $\tau/\tau_\eta$ measured from DNS trajectories, for both full
  length trajectories (and with periodic boundary conditions) and for
  trajectories in smaller measurement volumes ${\cal
    L}/\mathcal{B}=1/4,1/7$. (b) The logarithmic local slope
  $\zeta_4(\tau)$ measured from DNS trajectories, for both the full
  length trajectories (periodic boundary conditions) and for
  trajectories in smaller measurement volumes ${\cal
    L}/\mathcal{B}=1,1/2,1/4,1/7$. Note the tendency toward a less
  developed plateau, at smaller and smaller values, as the measurement
  volume decreases.}
\label{fig:ale}
\end{center}
\end{figure}
As noted above, the EXP4 data for $\zeta_4(\tau)$ develop an apparent
plateau at a smaller value than the DNS data. In this section, we show
how the DNS data can be used to suggest a possible origin for this
mismatch.

We investigate the behavior of the local slopes for the simulations,
when the volume of size ${\cal L}^3$, where particles are tracked, is
systematically decreased.  Essentially only trajectories which stay in
this sub-volume are considered in the analysis, mimicking what happens
in the experimental measurement volume.  We considered volume sizes
$\mathcal{L}$ in the range which goes from the full box size
$\mathcal{B}$ to $\mathcal{B}/7$, and we average over all the
sub-boxes to increase the statistical samples. In
Fig.~\ref{fig:times}, we plot the statistics of the trajectory
durations for both the experiment and DNS by varying the measurement
volume size. For ${\cal L}=\mathcal{B}/4$, the modified DNS statistics
are essentially indistinguishable from the experimental results.  It
is now interesting to look at the LVSF measured from these finite
length numerical trajectories.  This shows that the method we devised
is able to mimic the presence of a finite measurament volume as in
experiments.

In Fig.~\ref{fig:aleA}, we show the fourth-order LVSF obtained by
considering the full length trajectories and the trajectories living
in a sub-volume as explained above.  What clearly appears from
Fig.~\ref{fig:aleA} is that the finite length of the trajectories
lowers the value of the structure functions for time lags of the order
of $20 \tau_\eta \le \tau \le 40 \tau_{\eta}$.  Indeed, the
finite-length statistics give a signal that is always lower than the
full averaged quantity: this effect may be due to a bias to slow, less
energetic particles, which have a tendency to linger inside the volume
for longer times than fast particles, introducing a systematic change
in the statistics.  Note that this is the same trend detected when
comparing EXP2 and EXP4 in Figs.~\ref{fig:LSF}.  In
Fig.~\ref{fig:aleB}, we also show the effect of the finite measurement
volume on the local slope for $p=4$. By decreasing the observation
volume, we observe a trend towards a shorter and shorter plateau with
smaller and smaller values. This could be the source of the small
offset between the plateaux developed by the EXP4 data and the DNS
data in Fig.~\ref{fig:6}.

For the sake of clarity, we should recall that in the DNS particles
can travel across a cubic fully periodic volume, so during their full
history they can reenter the volume several times. In principle, this
may affect the results for long time delays.  However, since the
particle velocity is taken at different times we may expect that
possible spurious correlations induced by the periodicity to be very
small, if not absent. This is indeed confirmed in Fig.~\ref{fig:aleB}
where we can notice the perfect agreement between data obtained by
using periodic boundary conditions or limiting the analysis to
subvolumes of size $\mathcal{L}=\mathcal{B}$ (i.e. not retaining the
periodicity) and even $\mathcal{L}=\mathcal{B}/2$.

\subsubsection{Filtering and measurement error effects at small time lags}

\begin{figure}[t!]
\begin{center}
  \subfigure[]{
    \includegraphics[width=0.47\textwidth]{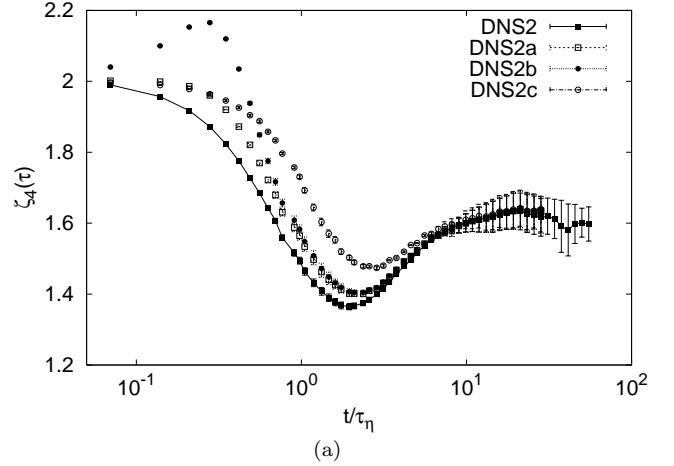}
    \label{fig:filterDNS}
  } \subfigure[]{
    \includegraphics[width=0.47\textwidth]{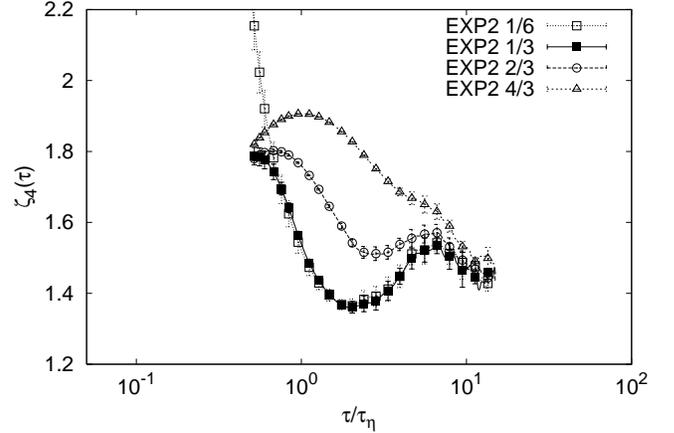}
\label{fig:filterEXP}
}
\caption{(a) Logarithmic local slope $\zeta_{4}(\tau)$ for the DNS2
  data set. The symbol DNS2a,b, and c denote the DNS2 trajectories
  modified by noise and filter effects, mimicking what was done in the
  experiments. In particular, DNS2a refers to the introduction of
  noise in the particle position of the order of $\delta x \sim
  \eta/10$ and with a Gaussian filter width $\sigma \sim
  \tau_{\eta}/3$, DNS2b to the same filter width but with much larger
  spatial noise ($\delta x \sim \eta/4$), and DNS2c to the same
  spatial noise but a large filter width $\sigma \sim 2
  \tau_{\eta}/3$.  Note how when the filter is not very large and with
  large spatial errors we have strong non-monotonic behavior for the
  local slopes (DNS2b). (b) The effect of filter width on data from
  EXP2 experiment ($R_\lambda = 690$, small measurement volume).  We
  tested 4 different filter widths: $\sigma / \tau_\eta$ = 1/6, 1/3,
  2/3, and 4/3. }
\label{fig:filter}
\end{center}
\end{figure}

As discussed in Sect.~\ref{sect:expdns}, results at small time lags
can be slightly contaminated by several effects both in DNS and
experiments.  DNS data can be biased by resolution effects due to
interpolation of the Eulerian velocity field at the particle position.
In experiments uncorrelated experimental noise needs to be filtered to
recover the trajectories.\cite{VSB98,VLCAB02,MCB04}

To understand the importance of such effects quantitatively, we have
modified the numerical Lagrangian trajectories in the following way. First, we
have introduced a random noise of the order of $\eta/10$ to the
particle position, in order to mimic the noise present in the
experimental particle detection.  Second, we have implemented the same
Gaussian filter of variable width used to smooth the experimental
trajectories ${\bm x}(t)$.  We also tested the effect of filtering by
processing experimental data with filters of different length.

In Figs.~\ref{fig:filterDNS} and \ref{fig:filterEXP}, we show the
local scaling exponents for $\zeta_4(\tau)$ as measured from these
modified DNS trajectories together with the results obtained from the
experiment, for several filter widths. The qualitative trend is very
similar for both the DNS and the experiment. The noise in particle
position introduces non-monotonic behavior in the local slopes at very
small time lags in the DNS trajectories.  This effect clearly
indicates that small scale noise may strongly perturb measurements at
small time lags, but will not have important consequences for the
behavior on time scales larger than $\tau_{\eta}$. On the other hand,
the effect of the filter is to increase the smoothness at small time
lags slightly (notice the shift of local slopes curves toward the
right for $\tau \sim \tau_{\eta}$ for increasing filter widths). A
similar trend is observed in the experimental data
(Fig.~\ref{fig:filterEXP}). In this case, choosing the filter width to
be in the range $\tau \in [1/6,1/3]\tau_{\eta}$ seems to be optimal,
minimizing the dependence on the filter width and the effects on the
relevant time lags. Understanding filter effects may be even more
important for experiments with larger particles, on the order of or
comparable with the Kolmogorov scale. In those cases, the particle
size naturally introduces a filtering by averaging velocity
fluctuations over its size, {\it i.e.}, those particles are not
faithfully following the fluid trajectories.\cite{VLCAB02,PINTON2}

\section{Conclusion and Perspectives}
\label{sec:concl}

A detailed comparison between state-of-the-art experimental and
numerical data of Lagrangian statistics in turbulent flows has been
presented. The focus has been on single-particle Lagrangian structure
functions.  Only due to the critical comparison of experimental and
DNS data it is possible to achieve a quantitative understanding of the
velocity scaling properties over the entire range of time scales, and
for a wide range of Reynolds numbers.

In particular, the availability of high Reynolds
number experimental measurements allowed us to assess in a robust way
the existence of very intense fluctuations, with high intermittency in
the Lagrangian statistics around $\tau \in [2:6] \tau_{\eta}$. For
larger time lags $\tau > 10 \tau_{\eta}$, the signature of different
statistics seems to emerge, with again good agreement between DNS and
experiment (see Fig.~\ref{fig:6}).  Whether the trend of logarithmic
local slopes at large times is becoming more and more extended at
larger and larger Reynolds number is an issue for further
research.

Both experiments and numerics show in the ESS local slope of the
fourth and sixth order Lagrangian structure functions a dip region at
around time lags $(2\div 6) \tau_{\eta}$ and a flattening at $\tau >
10 \tau_{\eta}$. As of today, it is unclear whether the dip or the
flattening region give the asymptotic scaling properties of Lagrangian
turbulence.  The question of which region will extend as a function of
Reynolds number can not be resolved at present, and remains open for
future research.

It would also be important to probe the possible relations between
Eulerian and Lagrangian statistics as suggested by simple
phenomenological multifractal
models.\cite{borgas,boffetta,BBCLT05,PINTON1} In these models, the
translation between Eulerian (single-time) spatial statistics and
Lagrangian statistics is made via the dimensional expression of the
local eddy turnover time at scale $r$: $\tau_r \sim r/\delta_r u$.
This allows predictions for Lagrangian statistics if the Eulerian
counterpart is known. An interesting application concerns Lagrangian
acceleration statistics,\cite{BBCLT05} where this procedure has given
excellent agreement with experimental measurements. When applied to
single-particle velocities, multifractal predictions for the LVSF
scaling exponents are close to the plateau values observed in DNS at
time lags $\tau > 10 \tau_{\eta}$. It is not at all clear, however, if
this formalism is able to capture the complex behavior of the local
scaling exponents close to the dip region $\tau \in [2:6]
\tau_{\eta}$, as depicted in Fig.~\ref{fig:6}. Indeed, multifractal
phenomenology, as with all multiplicative random cascade
models,\cite{frisch} does not contain any signature of spatial
structures such as vortex filaments. It is possible that in the
Lagrangian framework a more refined matching to the viscous
dissipative scaling is needed, as was proposed in
Ref.~\onlinecite{PINTON1}, rephrasing known results for Eulerian
statistics.\cite{fv} Even less clear is the relevance for Lagrangian
turbulence of other phenomenological models, based on
super-statistics~\cite{beck}, as recently questioned in
Ref.~\onlinecite{gotoh}.

The formulation of a stochastic model able to capture the whole shape
of local scaling properties from the smallest to the largest time lag,
as depicted in Fig.~\ref{fig:6}, remains an open important theoretical
challenge.

EB, NTO and HX gratefully acknowledge financial
support from the NSF under contract PHY-9988755 and PHY-0216406 and by
the Max Planck Society. LB, MC, ASL and FT acknowledge J. Bec,
G. Boffetta, A. Celani, B. J. Devenish and S. Musacchio for
discussions and collaboration in previous analysis of the numerical
dataset.  LB acknowledges partial support from MIUR under the project
PRIN 2006. Numerical simulations were performed at CINECA (Italy)
under the ``key-project'' grant: we thank G. Erbacci and C. Cavazzoni
for resources allocation. LB, MC, ASL and FT thank the DEISA
Consortium (co-funded by the EU, FP6 project 508830), for support
within the DEISA Extreme Computing Initiative (www.deisa.org).
Unprocessed numerical data used in this study are freely available
from the iCFDdatabase.\cite{iCFDdatabase}


\begin{thebibliography}{99}


\bibitem{P94} S. B. Pope, ``Lagrangian PDF methods for turbulent flows,'' 
Annu. Rev. Fluid Mech.~\textbf{26}, 23 (1994).

\bibitem{Pope} S. B. Pope, \textit{Turbulent Flows}, (Cambridge
  University Press, Cambridge UK, 2000)

\bibitem{S01} B. Sawford, ``Turbulent relative dispersion,''
  Annu. Rev. Fluid Mech.~\textbf{33}, 289 (2001).

\bibitem{Y02} P. K. Yeung, ``Lagrangian investigations of turbulence,''
Annu. Rev. Fluid Mech.~\textbf{34}, 115 (2002).

\bibitem{YP89} P. K. Yeung and S. B. Pope, ``Lagrangian statistics
  from direct numerical simulations of isotropic turbulence,''
  J. Fluid Mech.~\textbf{207}, 531 (1989).

\bibitem{VD97} M. Virant and Th. Dracos, ``3D PTV and its application
  on Lagrangian motion,'' Meas.~Sci.~Technol.~\textbf{8}, 1539 (1997).

\bibitem{VSB98} G. A. Voth, K. Satyanarayan and E. Bodenschatz, 
``Lagrangian acceleration measurements at large Reynolds numbers,'' 
Phys. Fluids.~\textbf{10}, 2268 (1998).

\bibitem{OM00} S. Ott and J. Mann, ``An experimental investigation of
  the relative diffusion of particle pairs in three-dimensional
  turbulent flow,'' J. Fluid Mech.~\textbf{422}, 207 (2000).

\bibitem{LVCAB01} A. La Porta, G. A. Voth, A. M. Crawford,
  J. Alexander and E. Bodenschatz, ``Fluid particle accelerations in
  fully developed turbulence,'' Nature~\textbf{409}, 1017 (2001).

\bibitem{MMMP01} N. Mordant, P. Metz, O. Michel and J. F. Pinton,
  ``Measurement of Lagrangian velocity in fully developed
  turbulence,'' Phys. Rev. Lett.~\textbf{87}, 214501 (2001).

\bibitem{VLCAB02} G. A. Voth, A. La Porta, A. M.  Crawford,
  J. Alexander and E. Bodenschatz, ``Measurement of particle
  accelerations in fully developed turbulence,'' J. Fluid
  Mech.~\textbf{ 469}, 121 (2002).

\bibitem{SYBVLCB03} B. L. Sawford, P. K. Yeung, M. S. Borgas,
  P. Vedula, A. La Porta, A. M. Crawford, and E. Bodenschatz,
  ``Conditional and unconditional acceleration statistics in
  turbulence,'' Phys. Fluids~\textbf{15}, 3478 (2003).

\bibitem{PINTON1} L. Chevillard, S.G. Roux, E. Leveque, N. Mordant,
  J.-F. Pinton and A. Arneodo, ``Lagrangian velocity statistics in
  turbulent flows: Effects of dissipation,''
  Phys. Rev. Lett.~\textbf{91}, 214502 (2003).

\bibitem{mlt2003} I. M. Mazzitelli, D. Lohse and F. Toschi, ``Effect
  of microbubbles on developed turbulence,'' Phys. Fluids~\textbf{15},
  L5 (2003).

\bibitem{PINTON2} N. Mordant, E. Leveque and J.-F.  Pinton,
  ``Experimental and numerical study of the Lagrangian dynamics of
  high Reynolds turbulence,'' New J. Phys.~\textbf{6}, 116 (2004).

\bibitem{MCB04} N. Mordant, A.M. Crawford and E. Bodenschatz,
  ``Experimental Lagrangian acceleration probability density function
  measurement,'' Physica D~\textbf{193}, 245 (2004).

\bibitem{ml2004} I. M. Mazzitelli and D. Lohse, ``Lagrangian
  statistics for fluid particles and bubbles in turbulence,'' New
  J. Phys.~\textbf{6}, 203 (2004).

\bibitem{BBCDLT04} L. Biferale, G. Boffetta, A. Celani, B. J. Devenish,
  A. Lanotte and F. Toschi, ``Multifractal statistics of Lagrangian
  velocity and acceleration in turbulence,''
  Phys. Rev. Lett.~\textbf{93}, 064502 (2004).

\bibitem{PK} P. K. Yeung, D. A. Donzis and K. R. Sreenivasan,
  ``High-Reynolds-number simulation of turbulent mixing,''
  Phys. Fluids~\textbf{17}, 081703 (2005).

\bibitem{luthi}B. Luthi, A. Tsinober, W. Kinzelbach, 
``Lagrangian measurement of vorticity dynamics in turbulent flow,''
J. Fluid Mech.~\textbf{528}, 87 (2005)

\bibitem{leveque} L. Chevillard, SG. Roux, E. Leveque, N. Mordant,
J.-F. Pinton, A. Arneodo, ``Intermittency of velocity time increments
in turbulence,'' Phys. Rev. Lett.~\textbf{95}, 064501 (2005)

\bibitem{liberzon2} K. Hoyer, M. Holzner, B. Luthi, M. Guala,
A. Liberzon, W. Kinzelbach, ``3D scanning particle tracking
velocimetry,'' Exp. Fluids~\textbf{39}, 923 (2005)

\bibitem{BBCLT05} L. Biferale, G. Boffetta, A. Celani, A. Lanotte and
  F. Toschi, ``Particle trapping in three dimensional fully developed
  turbulence,'' Phys. Fluids.~\textbf{17}, 021701 (2005).

\bibitem{OXBB06b} N.T. Ouellette, H. Xu, M. Bourgoin, and
  E. Bodenschatz, ``Small-scale anisotropy in Lagrangian turbulence,''
  New J. Phys.~\textbf{8}, 102 (2006).

\bibitem{BBBCCLMT06} J. Bec, L. Biferale, G. Boffetta, A. Celani,
  M. Cencini, A. Lanotte, S. Musacchio, and F. Toschi, ``Acceleration
  statistics of heavy particles in turbulence,'' J. Fluid
  Mech.~\textbf{550}, 349 (2006).

\bibitem{X06} H. Xu, M. Bourgoin, N.T. Ouellette and E. Bodenschatz, 
``High order Lagrangian velocity statistics in turbulence,''
Phys. Rev. Lett.~\textbf{96}, 024503 (2006).

\bibitem{YPS06} P. K. Yeung, S. B. Pope and B. L. Sawford, ``Reynolds
  number dependence of Lagrangian statistics in large numerical
  simulations of isotropic turbulence,'' J. Turbul.~\textbf{7}, 58
  (2006).

\bibitem{OXBB06} N.T. Ouellette, H. Xu, M. Bourgoin and E. Bodenschatz, 
``An experimental study of turbulent relative dispersion models,''
New J. Phys.~\textbf{8}, 109 (2006).

\bibitem{BEC06} J. Bec, L. Biferale, M. Cencini, A. Lanotte and
  F. Toschi, ``Effects of vortex filaments on the velocity of tracers
  and heavy particle in turbulence,'' Phys. Fluids~\textbf{18}, 081702
  (2006).

\bibitem{berg} J. Berg, ``Lagrangian one-particle velocity statistics
in a turbulent flow,'' preprint arXiv:physics/0610155

\bibitem{liberzon1} M. Guala, A. Liberzon, A. Tsinober, W. Kinzelbach,
``An experimental investigation on Lagrangian correlations of
small-scale turbulence at low Reynolds number,'' J. Fluid
Mech.~\textbf{574}, 405 (2007).

\bibitem{AGCBW07} S. Ayyalasomayajula, A. Gylfason, L. R. Collins,
  E. Bodenschatz, Z. Warhaft, ``Lagrangian Measurements of Inertial
  Particle Accelerations in Grid Generated Wind Tunnel Turbulence,''
  Phys. Rev. Lett.~\textbf{97}, 144507 (2007).

\bibitem{YPKL07} P. K. Yeung, S. B. Pope, E. A. Kurth and A. G.
  Lamorgese, ``Lagrangian conditional statistics, acceleration and
  local relative motion in numerically simulated isotropic
  turbulence,'' J. Fluid. Mech.~\textbf{582}, 399 (2007).

\bibitem{MUELLER06} H. Homann, R. Grauer, A. Busse and W. C. M\"uller,
  ``Lagrangian Statistics of Navier-Stokes- and MHD-Turbulence,''
  preprint arXiv:physics/0702115v1

\bibitem{frisch} U. Frisch, \textit{Turbulence: the legacy of
    A.N. Kolmogorov} (Cambridge University Press, Cambridge UK, 1995)

\bibitem{rbm} R. Benzi, G. Paladin, G. Parisi and A. Vulpiani, ``On
  the multifractal nature of fully developed turbulence and chaotic
  systems,'' J. Phys. A~\textbf{17}, 3521 (1984).

\bibitem{sheleveque} Z. S. She and E. Leveque, ``Universal scaling
  laws in fully developed turbulence,'' Phys. Rev. Lett. \textbf{72},
  336 (1994).

\bibitem{castaign} B. Castaing, Y. Gagne and E. J. Hopfinger,
  ``Velocity probability density functions of high Reynolds number
  turbulence,'' Physica D~\textbf{46}, 435 (1990)

\bibitem{yakhot} V. Yakhot, ``Mean-field approximation and a small
  parameter in turbulence theory,'' Phys. Rev. E \textbf{63}, 026307
  (2001).

\bibitem{prlgradients} R. Benzi, L. Biferale, G. Paladin, A. Vulpiani
  and M. Vergassola, ``Multifractality in the statistics of the
  velocity gradients in turbulence,'' Phys. Rev. Lett.~\textbf{67},
  2299 (1991).
    
\bibitem{fv} U. Frisch and M. Vergassola, ``A prediction of the
  multifractal model -- The intermediate dissipation range,''
 Europhys. Lett.~\textbf{14}, 439 (1991).

\bibitem{arneoodo} A. Arneodo et al.  ``Structure functions in
  turbulence, in various flow configurations, at Reynolds number
  between 30 and 5000, using extended self-similarity,''
  Europhys. Lett.~\textbf{34}, 411 (1996).

\bibitem{SCHMIDT} F. G. Schmitt, ``Relating Lagrangian passive scalar
  scaling exponents to Eulerian scaling exponents in turbulence,''
  Physica A~\textbf{48}, 129 (2005).

\bibitem{beck} C. Beck, ``Statistics of three-dimensional Lagrangian
  turbulence,'' Phys. Rev. Lett.~\textbf{98}, 064502 (2007).

\bibitem{MY75} A. S. Monin and A. M. Yaglom, {\it Statistical Fluid 
Mechanics}, Vol. II, (MIT Press, Cambridge, MA, 1975).

 \bibitem{LD02} R.-C. Lien and E. A. D\'Asaro, ``The Kolmogorov
   constant for the Lagrangian spectrum and structure function,''
   Phys. Fluids~\textbf{14}, 4456 (2002).

\bibitem{borgas} M. S. Borgas, ``The Multifractal Lagrangian Nature of
  Turbulence,'' Phil. Trans. R. Soc. London A~\textbf{342}, 379
  (1993).

\bibitem{boffetta} G. Boffetta, F. De Lillo and S. Musacchio,
  ``Lagrangian statistics and temporal intermittency in a shell model
  of turbulence,'' Phys. Rev. E~\textbf{66}, 066307 (2002).

\bibitem{OXB06} N.T. Ouellette, H. Xu and E. Bodenschatz, ``A
quantitative study of three-dimensional Lagrangian particle tracking
algorithms,'' Exp. Fluids~\textbf{40}, 301 (2006).

\bibitem{XB07} H. Xu and E. Bodenschatz, ``Tracking Lagrangian
  trajectories in physical-velocity space,'' submitted (2007).

\bibitem{G02} T. Gotoh, D. Fukayama and T. Nakano, ``Velocity field
statistics in homogeneous steady turbulence obtained using a
high-resolution direct numerical simulation,'' 
Phys. Fluids~\textbf{14}, 1065 (2002).

\bibitem{KAN03} Y. Kaneda, T. Ishihara, M. Yokokawa, K. Itakura and
  A. Uno, ``Energy dissipation rate and energy spectrum in high
  resolution direct numerical simulations of turbulence in a periodic
  box,'' Phys. Fluids~\textbf{15}, L21 (2003).

\bibitem{shek} S. Chen, G. D. Doolen, R. H. Kraichnan and Z.-S. She, 
  ``On statistical correlations between velocity increments and
  locally averaged dissipation in homogeneous turbulence,''
  Phys. Fluids A~\textbf{5}, 458 (1993).

\bibitem{pk_spline} P. K. Yeung and S.B. Pope, ``An algorithm for
tracking fluid particles in numerical simulations of homogeneous
turbulence,'' J. Comput. Phys.~\textbf{79}, 373 (1988).

\bibitem{rovelstad_inter} A. L. Rovelstad, R. A. Handler and
  P. S. Bernard, ``The effect of interpolation errors on the
  Lagrangian analysis of simulated turbulent channel flow,''
  J. Comput. Phys.~\textbf{ 110}, 190 (1994).

\bibitem{tedeschi} H. Homann, J. Dreher and R. Grauer, ``Impact of the
  floating-point precision and interpolation scheme on the results of
  DNS of turbulence by pseudo-spectral codes,'' e-arXiv:0705.3144 to
  appear in Comp. Phys. Comm.

\bibitem{yakhot+sreene} V. Yakhot and K. R. Sreenivasan, ``Anomalous
  scaling of structure functions and dynamic constraints on turbulence
  simulations,'' J. Stat. Phys. \textbf{121}, 823 (2005).

\bibitem{bp} L. Biferale and I. Procaccia, ``Anisotropy in turbulent
  flows and in turbulent transport,'' Phys. Rep.~\textbf{414} 43,
  (2005).

\bibitem{ESS} R. Benzi, S. Ciliberto, R. Tripiccione, C. Baudet,
F. Massaioli and S. Succi, ``Extended self-similarity in turbulent
flows,'' Phys. Rev. E~\textbf{48}, R29 (1993)

\bibitem{gotoh} T. Gotoh and R. H. Kraichnan, ``Turbulence and Tsallis
  statistics,'' Physica D~\textbf{193}, 231 (2004).

\bibitem{iCFDdatabase} iCFDdatabase \url{http://cfd.cineca.it.}

\end{thebibliography}
\end{document}